\documentclass[twocolumn]{emulateapj}

\usepackage{makecell}
\usepackage{amsmath}
\usepackage[utf8]{inputenc}
\usepackage{graphicx}
\usepackage{color}
\usepackage[colorlinks]{hyperref}
\hypersetup{
    colorlinks,	
    citecolor=blue,
}

\newcommand{\be}{\begin{eqnarray}}
\newcommand{\ee}{\end{eqnarray}}

\newcommand{\src}{GRS~1915+105}
\newcommand{\hxmt}{\textit{Insight}--HXMT}

\shorttitle{QPOs of GRS~1915+105}
\shortauthors{Liu et al.}

\begin{document}

\title{Testing evolution of LFQPOs with mass accretion rate in GRS 1915+105 with \textit{Insight}--HXMT}

\author{Honghui Liu\altaffilmark{1}, Long Ji\altaffilmark{2}, Cosimo~Bambi\altaffilmark{1,\dag}, Pankaj Jain\altaffilmark{3}, Ranjeev Misra\altaffilmark{4}, Divya Rawat\altaffilmark{3}, J~S~Yadav\altaffilmark{3} and Yuexin Zhang\altaffilmark{5} }

\altaffiltext{1}{Center for Field Theory and Particle Physics and Department of Physics, 
Fudan University, 200438 Shanghai, China. \email[\dag E-mail: ]{bambi@fudan.edu.cn}} 
\altaffiltext{2}{Institut f\"ur Astronomie und Astrophysik, Kepler Center for Astro and Particle Physics, Eberhard Karls Universit\"at, T\"ubingen, Germany}
\altaffiltext{3}{Department of physics, IIT Kanpur, Kanpur, Uttar Pradesh 208016, India}
\altaffiltext{4}{Inter-University Center for Astronomy and Astrophysics, Ganeshkhind, Pune 411007, India}
\altaffiltext{5}{Kapteyn Astronomical Institute, University of Groningen, PO BOX 800, Groningen NL-9700 AV, the Netherlands}

\begin{abstract}
Using the \hxmt{} observations of \src{} when it exhibits low frequency quasi-periodic oscillations (QPOs), we measure the evolution of the QPO frequency along with disk inner radius and mass accretion rate. We find a tight positive correlation between the QPO frequency and mass accretion rate. Our results extend the finding of previous work with \textit{AstroSat} to a larger range of accretion rate with independent instruments and observations. Treating the QPO frequency of \src{} as the relativistic dynamic frequency of a truncated disk, we are able to confirm the high spin nature of the black hole in \src{}. We also address the potential of our finding to test general relativity in the future.
\end{abstract}



\section{Introduction}


Quasi-periodic oscillations \citep[QPOs;][]{Klis2005} in the form of narrow peaks in the power spectral density (PSD) are often observed in X-ray binaries (XRBs). The phenomenon is found to have similar characteristics in both neutron star and black hole accreting systems~\citep[e.g.][]{Wijnands1999}, suggesting a common physical origin (e.g. accretion/ejection flow). 

Over the last three decades, we have accumulated abundant knowledge about the behavior of QPOs in black hole XRBs. Correlation between the QPO frequencies and disk flux in hard state has been found in some systems~\citep{Remillard2006, Motta2011}. There is also clear evidence showing positive correlation between the lower frequency break ($\nu_{\rm b}$) and Low frequency QPOs (LFQPOs)~\citep[e.g.][]{Wijnands1999, Belloni2002}. Another tight correlation between LFQPOs and high frequency QPOs (or broad noise component) is found to exist over a large frequency range ~\citep{Psaltis1999}. These findings help to put strong constraints on models to explain QPOs.

LFQPOs have frequency roughly in the range 0.05--30 Hz. In black hole systems, LFQPOs can be divided into several types~\citep[Type A, B and C; See][]{Casella2004,Casella2005}. It has been found in some systems that different types of QPOs occur in different stages of the outburst of black hole transients~\citep{Motta2011}. Understanding the origin and mechanism of QPOs can give us important hints on both accretion/ejection process and the spacetime property near compact objects. For instance, \cite{Motta2014} {measured} the black hole spin of XTE~J1550-564 using a simultaneous detection of type-C and high frequency QPOs. There is also attempt to estimate black hole mass assuming that the correlation between LFQPOs and high energy spectrum index scales with mass~\citep{Shaposhnikov2007}.

Models proposed to explain LFQPOs refer mainly to instabilities or geometric effects. For instance, \cite{Tagger1999} proposed a model in which accretion-ejection instability (AEI) of a magnetized accretion disk can connect to QPOs observed in XRBs. \cite{Stella1998} interpreted the QPOs in low mass XRBs as Lense-Thirring precession of the innermost region of the accretion disk. Also in the frame of Lense-Thirring precession, \cite{Ingram2009} considered the precession of the hot flow inside a truncated disk and were able to explain why the observed maximum frequency is almost constant for all black hole XRBs. These models have been further extended and applied to observation data~\citep{Varniere2002, Titarchuk2004, Cabanac2010, Ingram2011, Veledina2013, Varniere2012, Karpouzas2020, Ma2020NA}. However, we have not reached a unified model that can explain all QPO behaviors.

Recently, \cite{Misra2020} confirmed the high spin of the black hole in \src{} using a spectral timing analysis of its X-ray radiation with \textit{AstroSat}~\citep{Yadav2016, Agrawal2017} data. The authors were able to simultaneously measure the QPO centroid frequency, disk inner radius and mass accretion rate. A correlation between the QPO frequency divided by the accretion rate and inner disk radius was found, which is expected if the QPO frequency is related to the dynamic frequency ($f_{\rm dyn}=c_{\rm s}(r)/r$, where $c_{\rm s}$ is the sound speed) of the standard accretion model. This kind of analysis requires both broadband energy coverage (for precision measurement of disk inner radius) and good timing ability of the instruments. We note that the Chinese X-ray satellite \textit{Hard X-ray Modulation Telescope} (dubbed as \hxmt{}, \cite{Zhang2014}) is also capable of measuring broadband energy spectrum and fast time variability from XRBs, which offers good opportunity to trace the coevolution of the QPOs and the disk parameters.

In this paper, we present a spectral timing analysis of \src{} observed by \hxmt{}. The data reduction procedure is summarized in Section~\ref{obs}. We describe the timing and spectral analysis in Section~\ref{d-ana}. We show the results in Section~\ref{res} and discuss the finding in Section~\ref{s-dis}.

\begin{table}
    \centering
    \renewcommand\arraystretch{2.0}
    \caption{\hxmt{} observations of \src{} analyzed in this paper}
    \label{info-obs}
    
    \begin{tabular*}{0.45\textwidth}{@{\extracolsep{\fill}}ccc}
        \hline\hline
        obsID & Date & Exposure (s)  \\
        \hline
        P0101330005 & 20180407 & 1759\\ 
        \hline
        P0101330006 & 20180409 & 5383\\ 
        \hline
        P0101330008 & 20180411 & 898\\ 
        \hline
        P0101330010 & 20180413 & 1399\\ 
        \hline
        P0101330011 & 20180414 & 3004\\ 
        \hline
        P0101330012 & 20180415 & 4795\\ 
        \hline
        P0101330013 & 20180416 & 4489\\ 
        \hline
        P0101330016 & 20180430 & 3494\\ 
        \hline
        P0101330017 & 20180506 & 4713\\ 
        \hline
        P0101310006 & 20180527 & 5043\\ 
        \hline
        P0101310007 & 20180601 & 6457\\ 
        \hline

\end{tabular*}\\

\textit{Note}. Only LE exposures are listed. The observation date is presented in the form of \texttt{yyyymmdd}.
\end{table}


\section{Observation and data reduction}\label{obs}

\src{} is a special low mass X-ray binary (LMXB) discovered in 1992 by WATCH~\citep{Castro-Tirado1992}. Unlike other black hole LMXBs that spend most time in quiescence, the source has been a persistent system since its discovery. It does not follow the typical Q-shape pattern on the hardness intensity diagram, but exhibits much more complex variability instead~\citep[see][]{Belloni2000, Hannikainen2003}.

\begin{figure*}
    \centering
    \includegraphics[width=\linewidth]{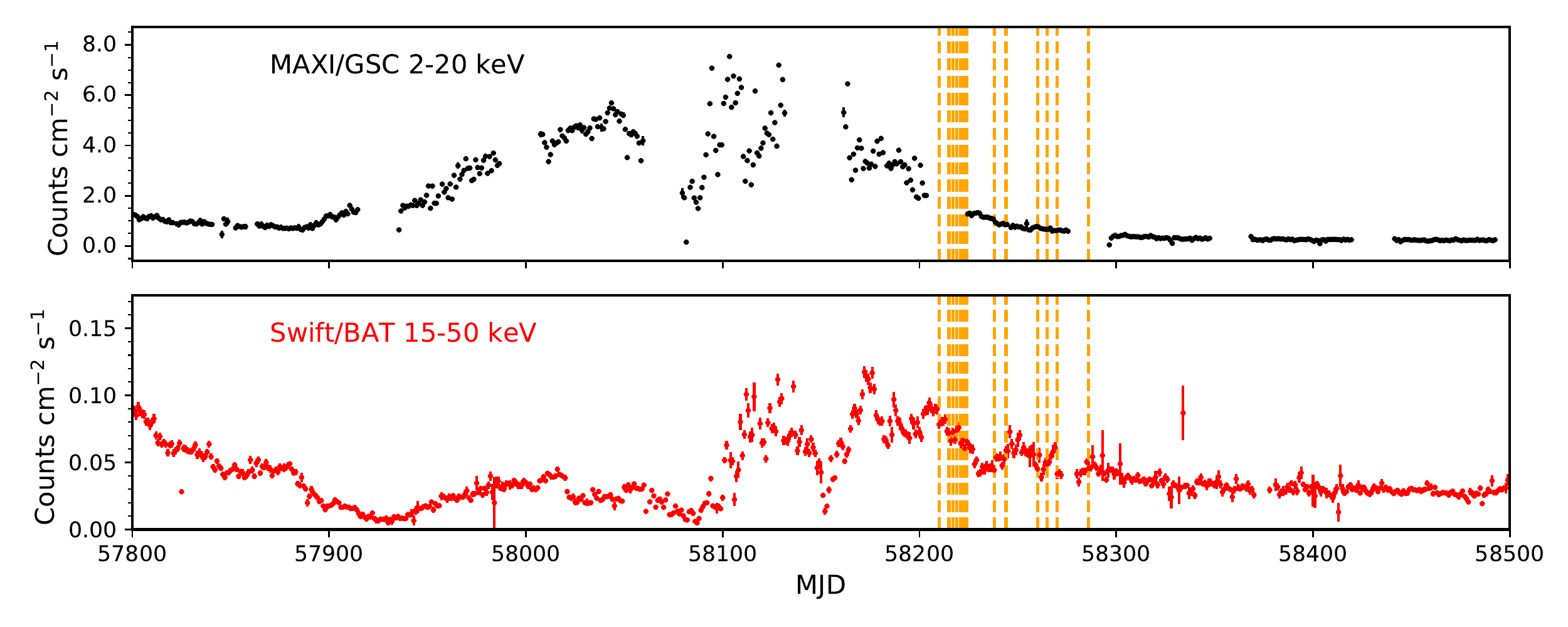}
    \caption{MAXI/GSC and Swift/BAT light curves of \src{} starting from February 2017. The vertical orange lines mark \hxmt\ observations.}
    \vspace{0.5cm}
    \label{lcurve}
\end{figure*}

\hxmt\ extensively observed \src\ from {2017 to 2020}. We went through all available \hxmt\ data of \src\ and {picked out those observations that show QPO signatures. We checked the lightcurve of each observation and excluded the ones with strong variability (e.g. strong flares or dips). Short exposures on the same day are combined after examing the stability of their lightcurves.} The selected observations analyzed in this work are marked in the lightcurve of \src{} in Figure~\ref{lcurve}. Information of these observations are listed in Table~\ref{info-obs}. We also show the hardness ratio and hardness intensity diagram of \src{} since 2009 in Figure~\ref{hid}. {We use the nearly daily monitoring data of \src{} from MAXI and Swift/BAT to create Figure~\ref{hid}. The count rates from both instruments are first scaled into Crab units and the hardness ratio is defined as the ratio between the scaled MAXI and Swift/BAT count rate.}

\textit{Insight}-HXMT is the first Chinese X-ray telescope, which consists of low-energy, medium-energy and high-energy detectors covering the broadband energy range of 1-250 keV \citep{Chen2020, Cao2020, Liu2020, Zhang2020}.  We extract lightcurves and spectra following the official user guide\footnote{\url{http://www.hxmt.cn/SoftDoc/67.jhtml}} and using the software {\sc HXMTDAS} ver 2.02. The background is estimated by standalone scripts \texttt{hebkgmap}, \texttt{mebkgmap} and \texttt{lebkgmap} \citep{Liao2020a, Guo2020, Liao2020b}. We screen good time intervals by considering the recommended criteria, i.e., the elevation angle $>$ 10 degree, the geomagnetic cutoff rigidity $>$ 8 GeV, the pointing offset angle $<$ 0.1 and at least 300 s away from the South Atlantic Anomaly (SAA).

We fit spectral data from \hxmt\ Low Energy X-ray Telescope (LE) in energy range 2--9 keV and Medium Energy X-ray Telescope (ME) in 8--20 keV. The High Energy X-ray Telescope (HE) data is not included because of strong background presence. We have also checked that adding HE data made little difference to the best-fit parameters.

\section{Data analysis} \label{d-ana}

\subsection{Timing analysis}

We extract the LE light curve of \src{} in the 1--10 keV energy band with a time resolution of 1/128 s. {Note that we use data in 2--9 keV for spectral analysis (instead of the 1--10 keV for timing analyis) because of calibration uncertainties of LE. The nominal time resolution of LE is 1 ms (corresponding to the Nyquist frequency of 500 Hz), but we are only interested in LFQPOs below 30 Hz in this study and thus a 1/128 s resolution is enough.}

We then use the \texttt{Python} package \texttt{Stingray}~\citep{Huppenkothen2019} to calculate the power spectral density (PSD) {with a segment size of 64 s}. {The final PSD is obtained by averaging all 64 s segments and is normalized according to~\cite{Belloni1990}}. We logarithmically rebin the PSD so that each bin size is 1.02 times larger than the previous bin. The PSD is then fitted in XSPEC between 0.1--20 Hz using several Lorentzian components~\citep{Belloni2002}. We need at least one narrow Lorentzian for the QPO and one zero-centered Lorentzian to fit broader component. More narrow Lorentzians are sometimes included to model harmonic peaks. All QPOs we detect have quality factor ($Q$) greater than 4 and detection significance greater than 3$\sigma$\footnote{The ratio of Lorentzian norm divided by its 1$\sigma$ negative error is larger than 3.}. One typical PSD is shown in Figure~\ref{psd_spec}. The QPO frequencies we find for each observation are listed in Table~\ref{bestfit}. {The ME lightcurve in 8--30 keV band has been analyzed in the same way and returns consistent measurements of the QPO frequencies. So we report only the results from LE data in Table~\ref{bestfit}.}

\subsection{Spectral analysis}

We use XSPEC v12.10.1f~\citep{xspec} with cross section set to~\cite{Verner1996} to analyze spectra of \src. As for element abundances, we test both~\cite{Anders1989} and~\cite{Wilms2000}. We find that the choice of abundances does not influence much the best-fit parameters (except for $n_{\rm H}$). This is consistent with what found by~\cite{Shreeram2020} in the reflection spectrum of \src. We therefore proceed 
with further analysis using abundances of~\cite{Wilms2000}, which is more up to date.

The \hxmt\ spectra of \src\ are fitted with model \texttt{Tbabs$\times$(simpl$\times$kerrd + kerrdisk)}. Model \texttt{Tbabs} accounts for absorption by Inter-Stellar Medium (ISM) and we set its column density ($n_{\rm H}$) to be free during the fitting. \texttt{kerrd}~\citep{Ebisawa2003} is included to model the emission from the optically thick accretion disk. The black hole mass, distance and inclination of the accretion disk are set to 12.4~$M_{\odot}$, 8.6~kpc and 60$^{\circ}$~\citep{Reid2014}, respectively. We also set the spectra hardening factor of \texttt{kerrd} to 1.7~\citep{Shimura1995}. Comptonization of disk photons is also taken into account by convolving \texttt{simpl}~\citep{Steiner2009} with \texttt{kerrd}. \texttt{kerrdisk}~\citep{Brenneman2006} is used to fit possible blurred fluorescent emission from the accretion disk. The rest frame line energy is fixed at 6.4~keV~\citep{Blum2009}. We fix the spin parameter ($a_*$) to 0.98 and index of the emissivity profile to 1.8 as did by~\cite{Misra2020}. Leaving these parameters free will not affect much the best-fit values of other parameters. {The disk reflection component is always weak (as shown in Figure~\ref{psd_spec}) in the analyzed observations, and the reflection can be well fitted with a simple guassian at iron band. So we do not consider more sophisticated reflection model (e.g. \texttt{relxill}~\citep{Garcia2014}).}

We run a Monte Carlo Markov Chain (MCMC) for each spectrum to estimate the uncertainties of free parameters. The XSPEC implementation of MCMC simulation (\texttt{Chain} command) using algorithm of~\cite{GW2010} is used to generate the chain. We set up 100 walkers to search the parameter space with the first 5000 steps ignored (burn in). The chain lengths differ from case to case, depending on the autocorrelation of each parameter, but we ensure each walker runs 30 times more steps than the longest autocorrelation length. To further test the convergence of the chain, we compare the two dimensional distribution for each pair of parameters from the first and second halves of the chain and we find no large difference.

\begin{figure}
    \centering
    \includegraphics[width=\linewidth]{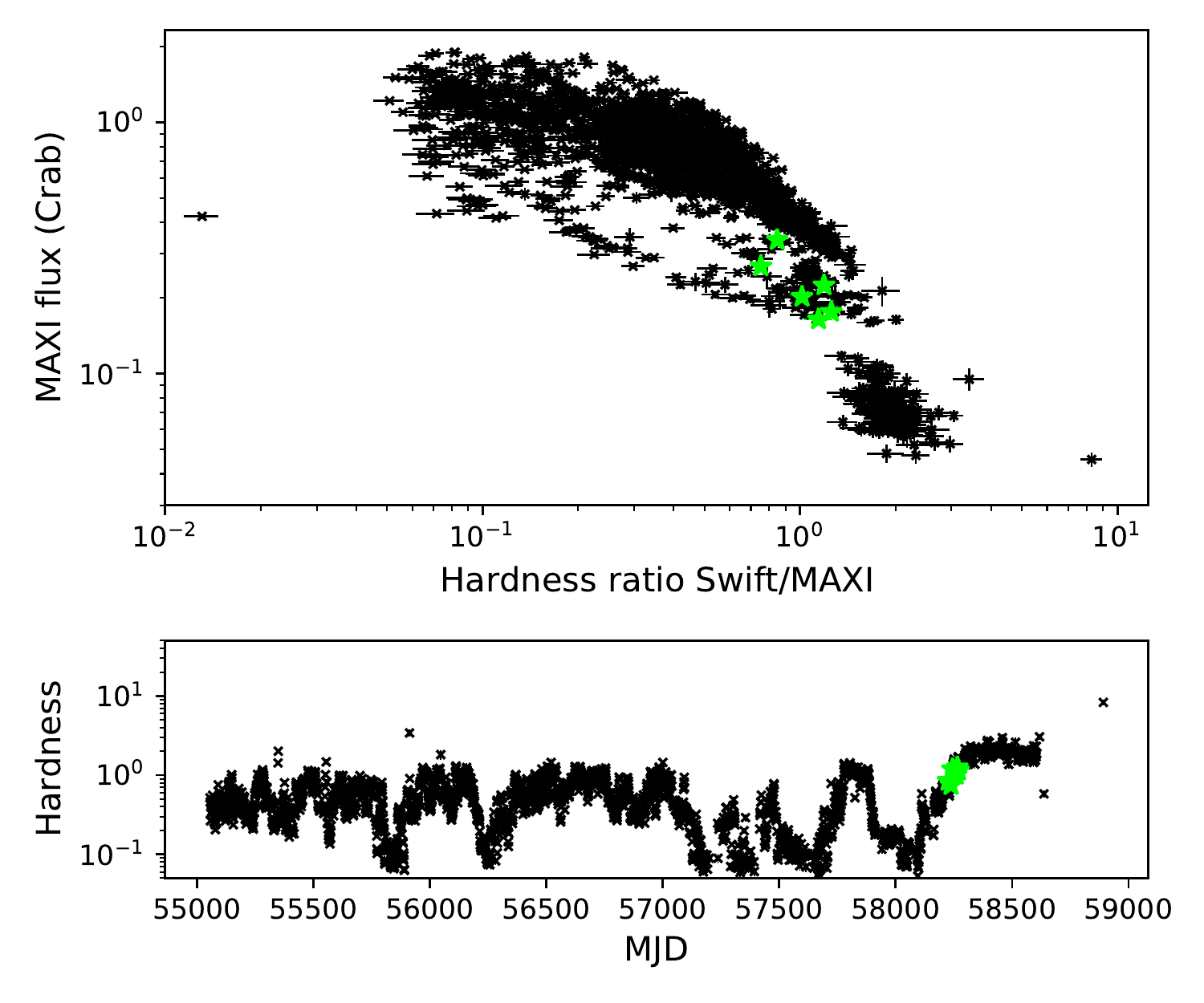}
    \caption{Hardness intensity diagram (HID) of \src\ from MAXI/GSC (2--20 keV) and Swift/BAT (15--50 keV) monitoring. Hardness ratio is defined as MAXI count rate in Crab units divided by Swift count rate. The \hxmt{} observations analyzed in this work are marked with green stars.}
    \label{hid}
\end{figure}

\begin{figure*}
    \centering
    \includegraphics[width=0.49\linewidth]{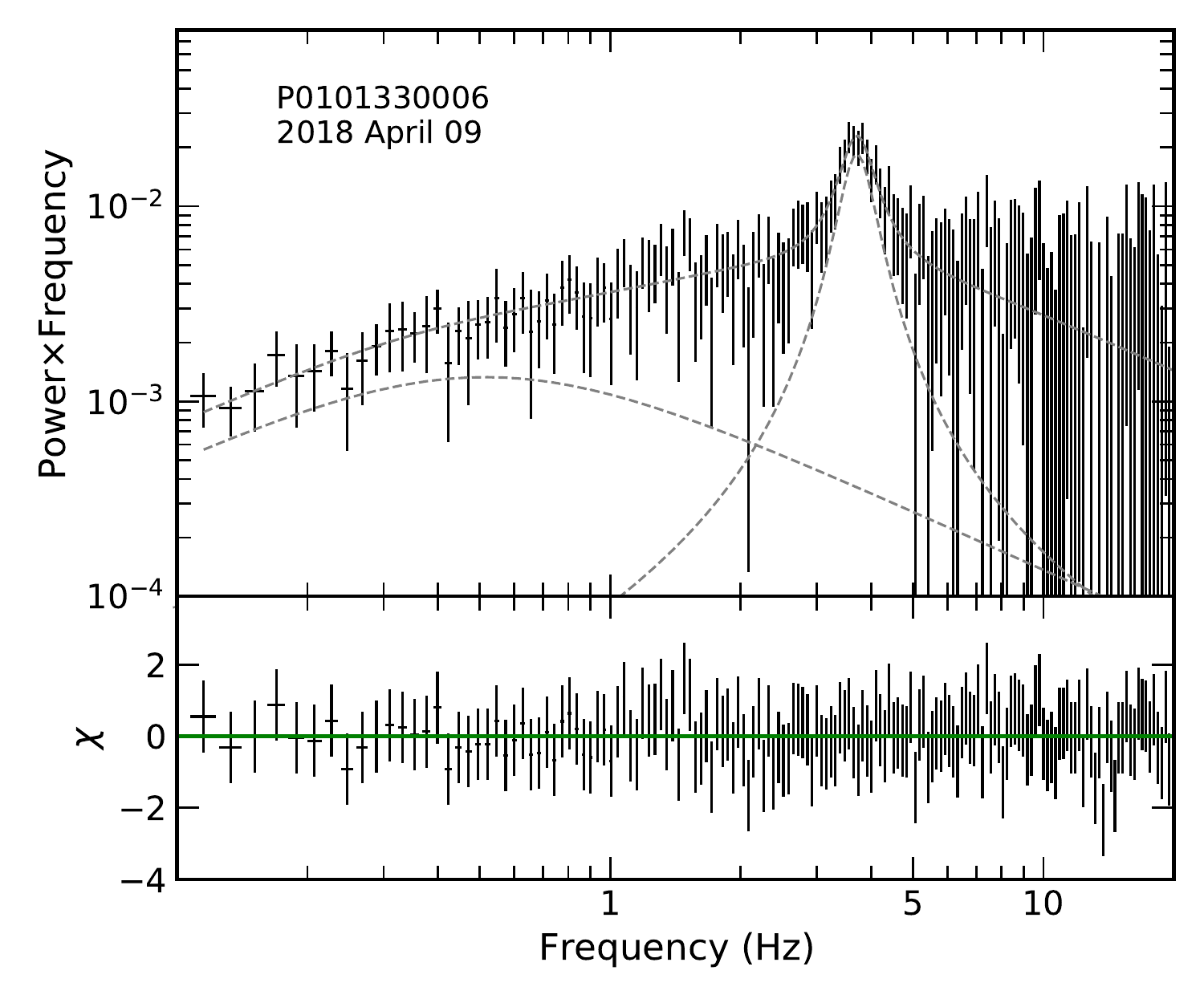}
    \includegraphics[width=0.49\linewidth]{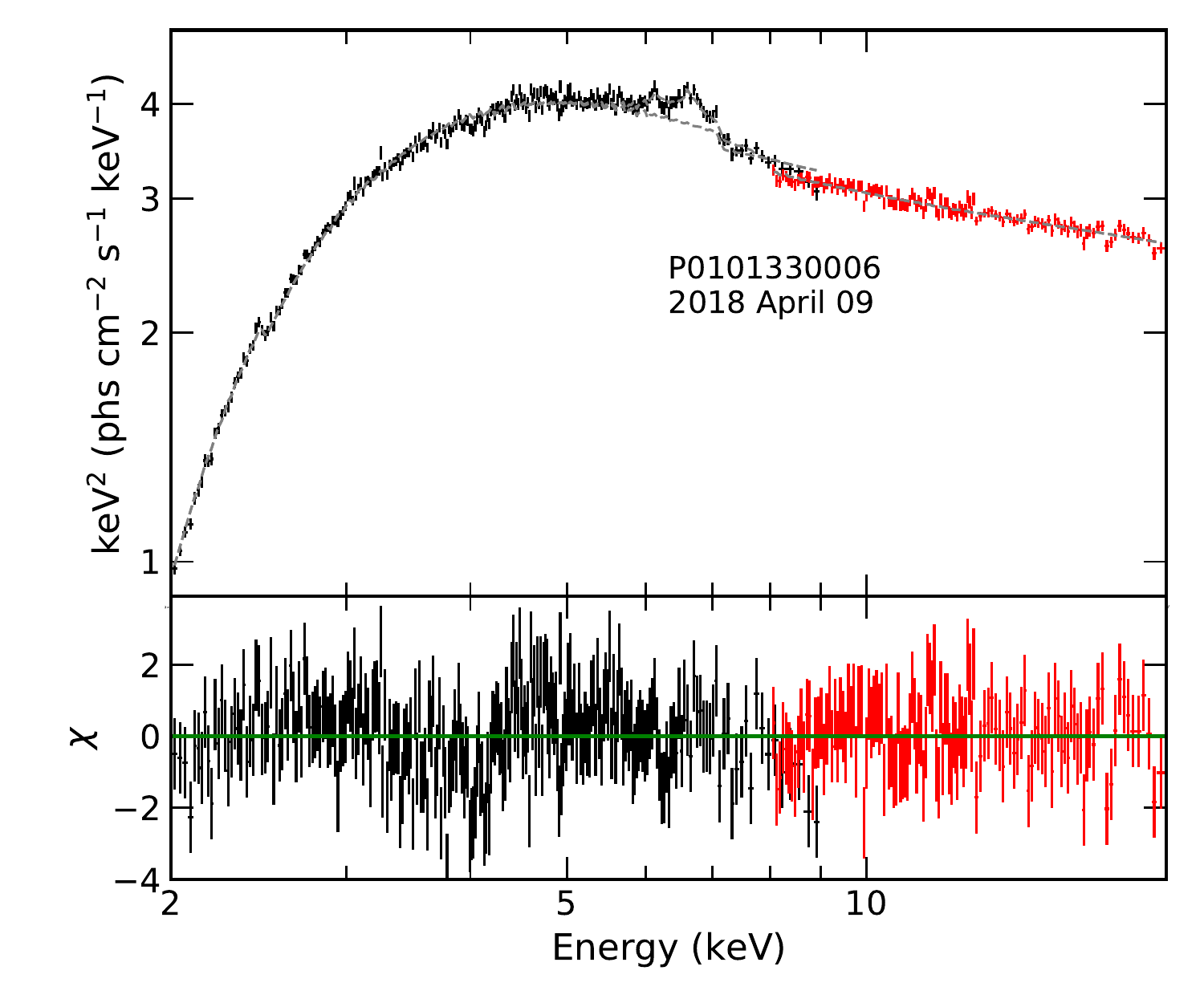}
    \caption{Left: The power spectral density (PSD) of \src\ observed by \hxmt\ on 2018 April 9 in 1--10~keV. Right: \hxmt\ spectrum of \src\ and residuals to the best-fit model. Data from LE and ME are plotted in black and red respectively.}
    \vspace{0.69cm}
    \label{psd_spec}
\end{figure*}

\section{results}\label{res}

The best-fit values and errors extracted from the chains are shown in Table~\ref{bestfit}. It is interesting to note that although the observations analyzed here spread over a 2-month interval and the source luminosity decreases by a factor of 3, the QPO signature is always clearly detected. {It might be due to that the corona-disk geometry does not change much during the interval, since the source hardness remains similar for the analyzed observations (see the bottom panel of Fugure~\ref{hid}).} 

\begin{figure*}
    \centering
    \includegraphics[width=\linewidth]{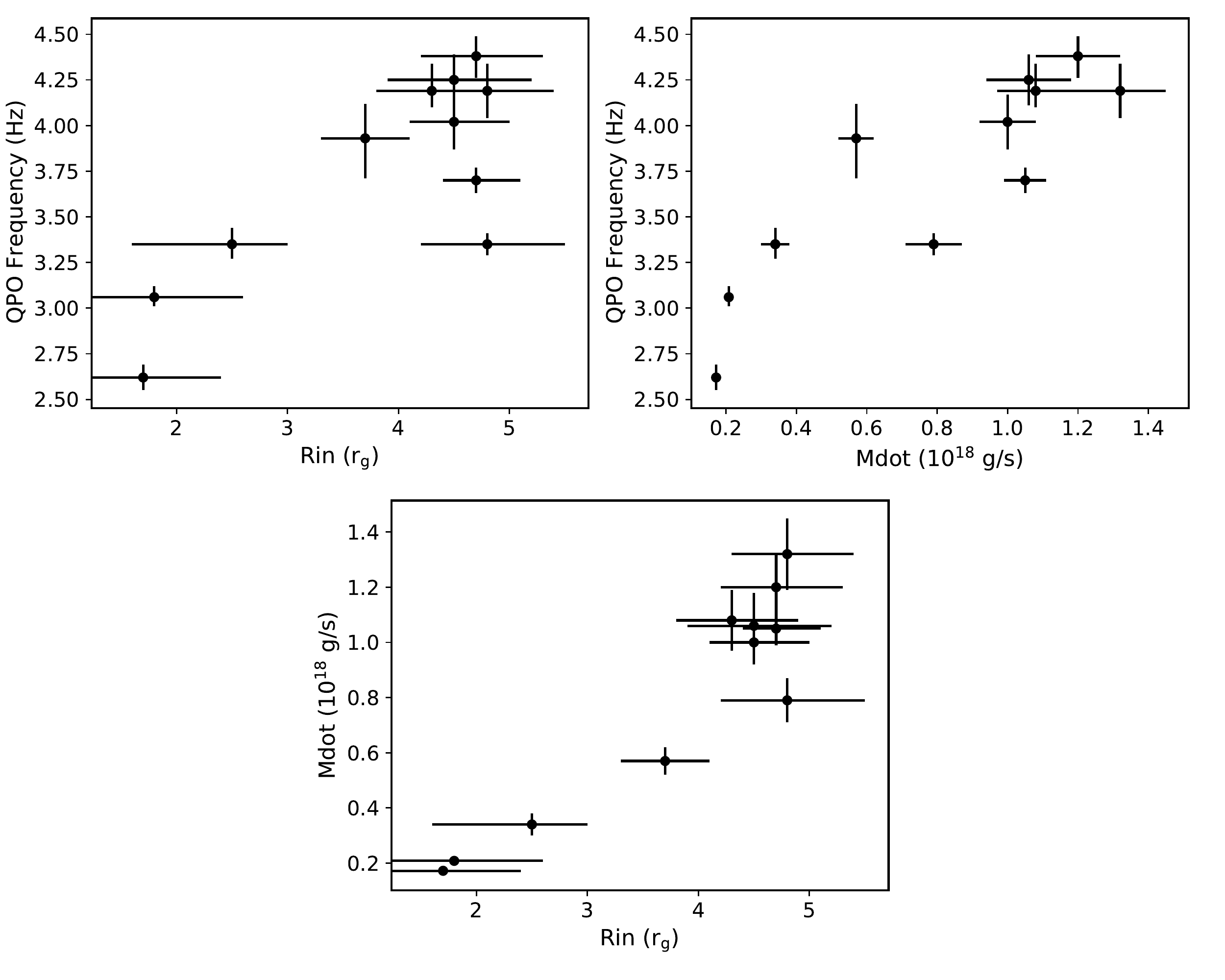}
    \caption{QPO frequency v.s. disk inner radius (upper left), QPO frequency v.s. accretion rate (upper right) and accretion rate v.s. disk inner radius (lower).}
    \label{three-varia}
\end{figure*}

Correlation between the QPO frequency, disk inner radius and accretion rate are shown in Figure~\ref{three-varia}. We run Spearman's rank correlation analysis and find that the strongest correlation is between the QPO frequency and mass accretion rate (correlation coefficient $\rho = 0.89$ with probability of random results $P<10^{-3}$). The correlation between QPO frequency and disk inner radius is less significant ($\rho = 0.54$, $P=8.8\%$). \cite{Misra2020} find a stronger correlation between QPO frequency divided by accretion rate and disk inner radius than the correlation between each pair of these parameters. {However, the same correlation we find is weaker than that between the QPO frequency and accretion rate, although the correlation is still strong ($\rho = -0.81$, $P=0.3\%$). We note that the range of mass accretion rate we explored is much larger than that by~\cite{Misra2020}. This larger range enables us to find out the direct dependence of QPO frequency on mass accretion rate.}


{In~\cite{Misra2020}, the authors identify the QPO frequency of \src{} as the dynamic frequency of a truncated disk. The dynamic frequency is defined as the ratio between the sound crossing velocity at the inner disk and the truncation radius ($f_{\rm dyn}\sim c_{\rm s}(r)/r$). Assuming a standard relativistic accretion disk~\citep{Novikov1973}, the dynamic frequency is a function of black hole spin ($a_*$), mass accretion rate ($\dot{M}$), truncation radius ($R_{\rm in}$) and an overall normalization factor ($N$). Since we can measure the QPO frequency, accretion rate and the truncation radius of \src{} with spectral timing analysis, it is possible to infer the spin parameter by fitting the correlation of these parameters (using equation (3) of~\cite{Misra2020}).}

\begin{figure*}
    \centering
    \includegraphics[width=0.49\linewidth]{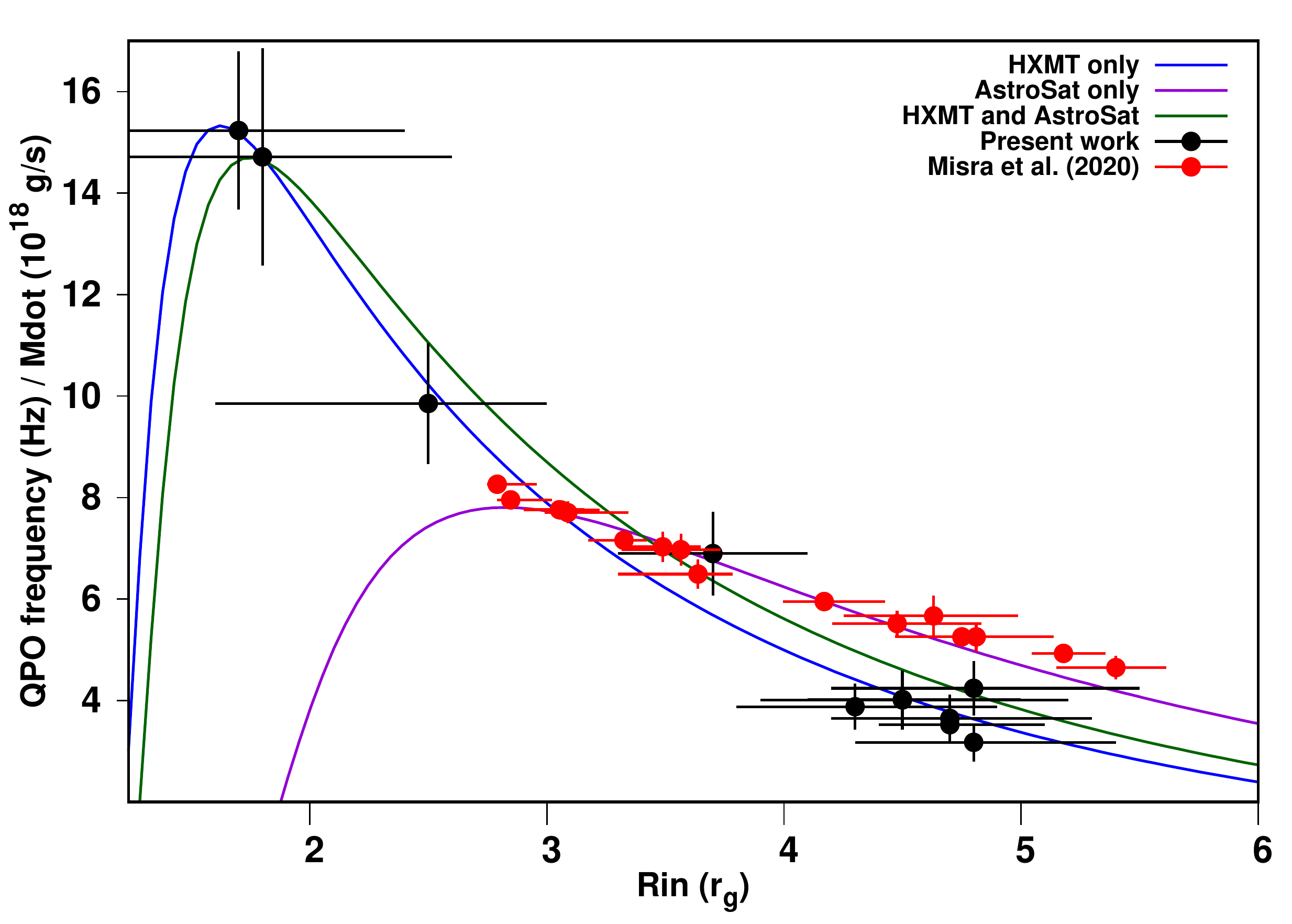}
    \caption{QPO frequency divided by accretion rate with disk inner radius. The black and red crosses denote results of this work and of \cite{Misra2020} respectively. The blue curve represents the best-fit of \hxmt{} data only (reduced $\chi^2=0.22$) and green curve for fitting both \hxmt{} and \textit{AstroSat} data (reduced $\chi^2=0.76$).}
    \label{relation}
\end{figure*}

We fit the relation between QPO frequency divided by accretion rate and disk inner radius using equation (3) of~\cite{Misra2020}. The fit on \hxmt{} data returns $a_* = 0.99897\pm 0.00019$ and $N = 0.108\pm 0.006$. We also try to fit the \hxmt{} and \textit{AstroSat} data simultaneously and we get $a_* = 0.99836\pm 0.00028$ and $N =  0.121\pm 0.005$, indicating a rapidly spinning black hole in \src{}. This measurement of high spin is consistent with what has been obtained by analyzing the blurred reflection spectra \citep{Blum2009, Miller2013, Zhang2019} or the thermal spectra \citep{McClintock2006} of \src{}. The best-fit curve is shown in Figure~\ref{relation}, as well as the results of~\cite{Misra2020}.

\begin{figure*}[htbp]
    \centering
    \includegraphics[width=0.49\linewidth]{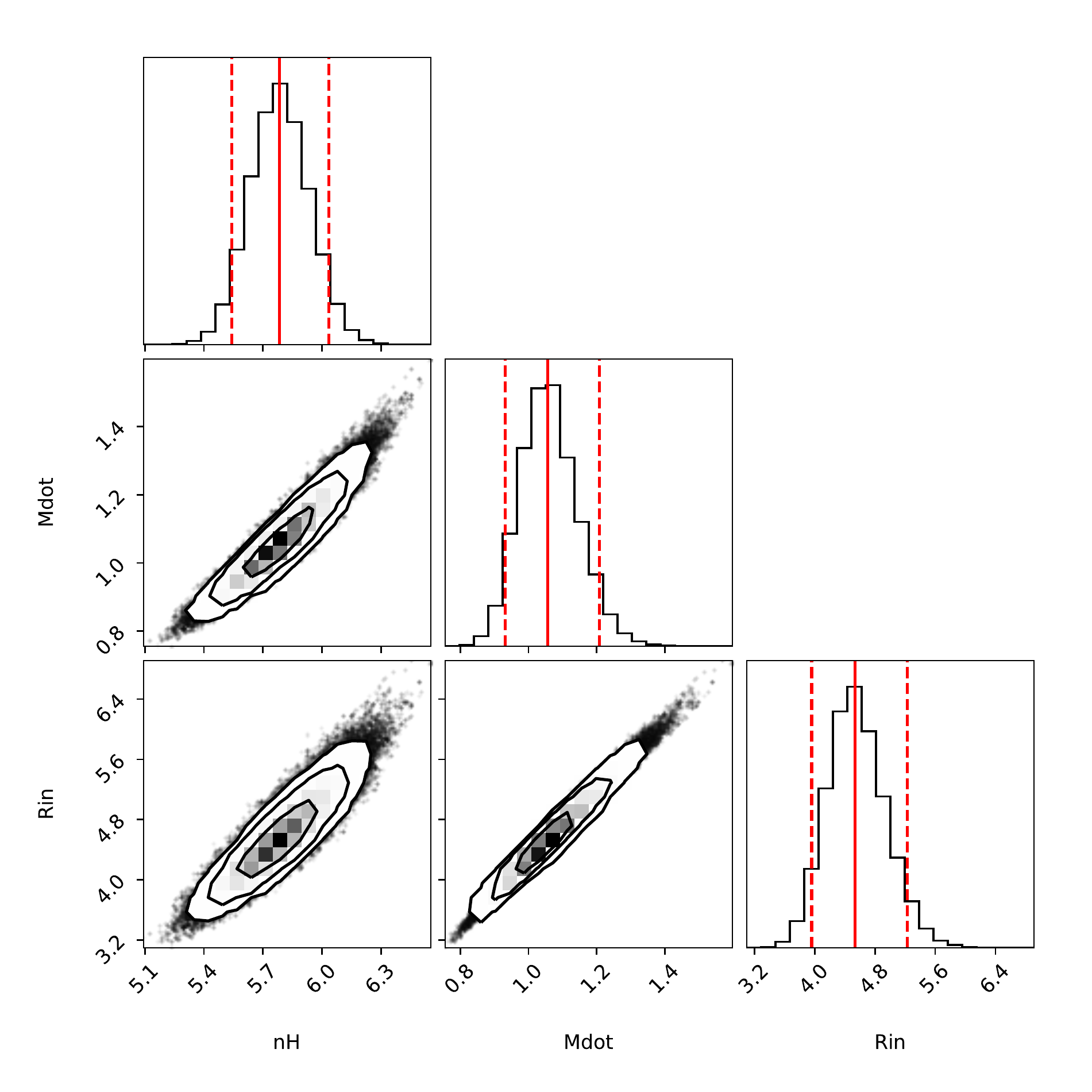}
    \includegraphics[width=0.49\linewidth]{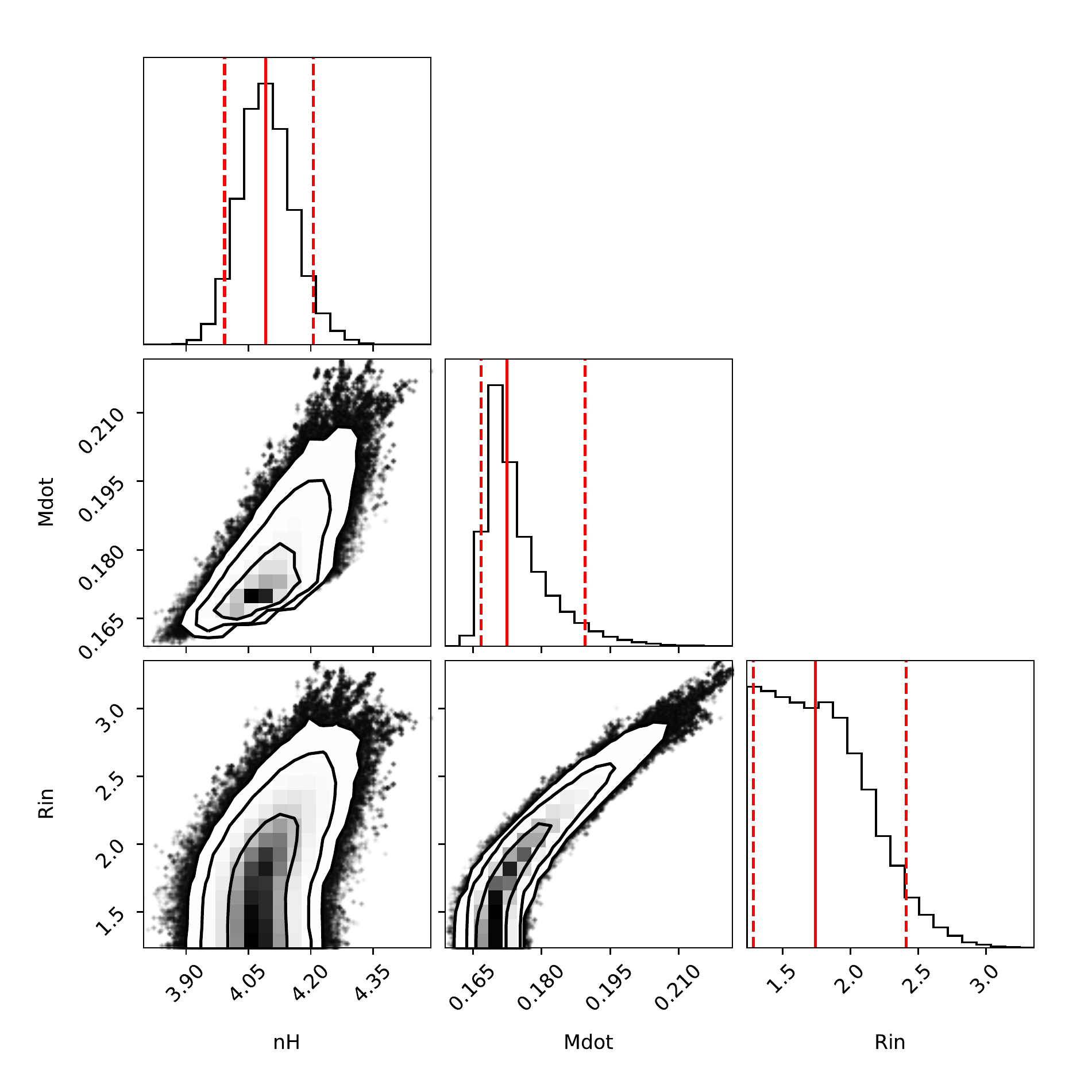}
    \caption{Correlation between the column density, disk inner radius and mass accretion rate for observations on 20180415 (left) and 20180601 (right). The red vertical lines denote the 5th, 50th and 95th percentiles for individual parameter.}
    \label{degeneracy}
\end{figure*}

From Table~\ref{bestfit}, we see that the column density of \texttt{Tbabs} evolves with inner radius and mass accretion rate, which raise the question if the degeneracy of these parameters affects the measurements. In Figure~\ref{degeneracy}, we plot degeneracy of the three parameters for two observations (one has the highest absorption column and the other has the lowest). There is indeed strong correlation between inner radius and mass accretion rate, which is expected since a smaller inner radius can somehow compensate the effect on the spectral shape by a lower accretion rate. This can also explain the degeneracy between column density of the absorption material and accretion rate. However, we find that these parameters are well constrained and we conclude letting nH free will not include bias on the measurement. We have also checked that holding the column density to be the same for all observations provides unacceptable fits. 

\begin{table*}
    \centering
    \renewcommand\arraystretch{2.0}
    \caption{Best fit parameters of GRS~1915+105}
    \label{bestfit}
    \begin{tabular}{cccccccccc}
        \hline\hline
        Date$^1$ & ${f_a}^2$ & ${f_u}^2$ & nH & Inner Radius & QPO frequency & Accretion Rate & $\Gamma$ & Fraction Scatter & $\chi^2$/Dof \\
        \hline
        &  &  &  ($10^{22}$ cm$^{-2}$) & ($R_{\rm g}$) & (Hz) & ($10^{18}$ g s$^{-1}$) & & & \\ \hline

        20180407 & 9.8 &13.8 & $5.66_{-0.2}^{+0.21}$ & $4.8_{-0.5}^{+0.6}$ & $4.19_{-0.15}^{+0.15}$ & $1.32_{-0.13}^{+0.15}$ & $2.32_{-0.05}^{+0.05}$ & $0.51_{-0.04}^{+0.05}$ & 436.53/426 \\ 
        \hline
        20180409 & 8.4 &11.4 & $5.27_{-0.11}^{+0.12}$ & $4.7_{-0.3}^{+0.4}$ & $3.7_{-0.07}^{+0.07}$ & $1.05_{-0.06}^{+0.08}$ & $2.23_{-0.022}^{+0.022}$ & $0.544_{-0.023}^{+0.025}$ & 845.44/811 \\ 
        \hline
        20180411 & 8.5 &12.0 & $5.61_{-0.22}^{+0.24}$ & $4.7_{-0.5}^{+0.6}$ & $4.38_{-0.12}^{+0.11}$ & $1.2_{-0.12}^{+0.15}$ & $2.28_{-0.04}^{+0.04}$ & $0.48_{-0.04}^{+0.04}$ & 341.95/373 \\ 
        \hline
        20180413 & 8.3 &11.7 & $5.6_{-0.2}^{+0.2}$ & $4.3_{-0.5}^{+0.6}$ & $4.19_{-0.09}^{+0.15}$ & $1.08_{-0.11}^{+0.12}$ & $2.22_{-0.06}^{+0.05}$ & $0.44_{-0.04}^{+0.04}$ & 329.53/388 \\ 
        \hline
        20180414 & 7.5 &10.6 & $5.56_{-0.16}^{+0.16}$ & $4.5_{-0.4}^{+0.5}$ & $4.02_{-0.15}^{+0.15}$ & $1.0_{-0.08}^{+0.09}$ & $2.26_{-0.04}^{+0.03}$ & $0.48_{-0.03}^{+0.03}$ & 608.07/635 \\ 
        \hline
        20180415 & 7.9 &11.0 & $5.78_{-0.24}^{+0.25}$ & $4.5_{-0.6}^{+0.7}$ & $4.25_{-0.14}^{+0.14}$ & $1.06_{-0.12}^{+0.15}$ & $2.22_{-0.06}^{+0.06}$ & $0.43_{-0.04}^{+0.05}$ & 301.74/331 \\ 
        \hline
        20180416 & 6.7 &9.1 & $5.1_{-0.17}^{+0.19}$ & $4.8_{-0.6}^{+0.7}$ & $3.35_{-0.06}^{+0.06}$ & $0.79_{-0.08}^{+0.1}$ & $2.15_{-0.022}^{+0.021}$ & $0.603_{-0.027}^{+0.029}$ & 833.9/743 \\ 
        \hline
        20180430 & 5.2 &7.1 & $5.06_{-0.16}^{+0.17}$ & $3.7_{-0.4}^{+0.4}$ & $3.93_{-0.22}^{+0.19}$ & $0.57_{-0.05}^{+0.06}$ & $2.175_{-0.027}^{+0.027}$ & $0.469_{-0.023}^{+0.024}$ & 589.93/616 \\ 
        \hline
        20180506 & 4.7 &6.1 & $4.35_{-0.18}^{+0.18}$ & $2.5_{-0.9}^{+0.5}$ & $3.35_{-0.08}^{+0.09}$ & $0.34_{-0.04}^{+0.04}$ & $2.08_{-0.023}^{+0.026}$ & $0.557_{-0.023}^{+0.027}$ & 708.31/652 \\ 
        \hline
        20180527 & 3.5 &4.5 & $4.3_{-0.1}^{+0.17}$ & {$1.8_{-P}^{+0.8}$} & $3.06_{-0.05}^{+0.06}$ & $0.208_{-0.008}^{+0.03}$ & $2.064_{-0.019}^{+0.02}$ & $0.655_{-0.027}^{+0.029}$ & 519.54/556 \\ 
        \hline
        20180601 & 3.3 &4.2 & $4.09_{-0.1}^{+0.12}$ & {$1.7_{-P}^{+0.7}$} & $2.62_{-0.07}^{+0.07}$ & $0.172_{-0.006}^{+0.017}$ & $2.05_{-0.016}^{+0.016}$ & $0.79_{-0.03}^{+0.03}$ & 669.21/621 \\ 
        \hline
        \hline
\end{tabular}\\

\textit{Note}. (1) The observation date is presented in the form of \texttt{yyyymmdd}. (2) The absorbed and unabsorbed flux in the 2--10 keV energy band in units 10$^{-9}$ ergs s$^{-1}$ cm$^{-2}$. Uncertainties correspond to the 5th and 95th percentiles from the MCMC samples. {The symbol $P$ means the errorbar touches the lower (or higher) limit.}
\end{table*}

\section{Discussion and conclusions}\label{s-dis}

With the advantage of broad energy coverage and good time resolution of \hxmt{}, we measure the evolution of LFQPOs signature of \src{} along with its mass accretion rate and inner disk radius. Assuming the QPO frequency corresponds to relativistic dynamic frequency of the disk, we are able to confirm the high spin nature of the black hole in \src{}. Our results extend the previous finding with \textit{AstroSat}~\citep{Misra2020}. 

{In Figure~\ref{relation}, the best-fit curve of~\cite{Misra2020} differs substantially from ours for low values of $R_{\rm in}$. This is due to the lack of data with low values of the disk inner radius in their analysis. At larger values of $R_{\rm in}$ ($R_{\rm in}>4 r_{\rm g}$ in Figure~\ref{relation}), our data is systematically lower than the \textit{AstroSat} result. This is why the joint fit (the green line) does not agree well with the data. We note that the dynamic frequency model is a very simple assumption. It might not be able to capture all the factors that drive the variability of LFQPOs of this source. So it is not a surprise that we find some differences between different observations.} 

{It is instructive to see that such a simple model can already roughly explain the behavior of LFQPOs in \src{}.} However, we note that there are still substantial locations in the hardness intensity diagram of the source not explored in this study (see Figure~\ref{hid}). Thus, more observations of \src{} by \hxmt{} (and \textit{AstroSat}) are certainly needed to understand the variability of this source.

We note that, although dynamic frequencies of a truncated disk can fit the behavior of LFQPOs in \src{}, an explanation of how the oscillation is generated is still missing. The Lense-Thirring precession model would predict direct dependence of QPO frequency on disk inner radius~\citep[See][]{Ingram2009, Ingram2010}. However, we find stronger dependence on mass accretion rate. The precession model also predicts anticorrelation between LFQPOs and disk inner radius, which is opposite to what we find (see upper left panel of Figure~\ref{three-varia}). These discrepancy may suggests that, in this particular case, the precession model is not favored. Note that in the ``transition layer model" proposed by~\cite{Titarchuk2004}, the mass accretion rate can indeed influence LFQPOs frequencies by affecting the $\gamma$-parameter (Reynolds number) of the accretion disk. The relation they derived between QPO frequency and $\gamma$-parameter (which is proportional to mass accretion rate) is very similar in shape to what we find (see Figure 3 of~\cite{Titarchuk2004} and upper right panel of Figure~\ref{three-varia}). This model might be promising to explain the origin of LFQPOs we find in \src{}, but further and more detailed investigation is certainly needed.

The lower panel of Figure~\ref{three-varia} shows that the inner disk radius tends to {decrease} when the accretion rate is decreasing (also when the source flux is decreasing). This is counterintuitive as we will expect the opposite for a truncated disk~\citep{Done2007}. {For typical X-ray transients (e.g. GX~339-4), the disk is believed to be truncated at large radii at the beginning of the outburst and the inner edge moves to smaller radii as the mass accretion rate increases~\citep[see][]{Esin1997}. There are evidences from observation that support this scenario~\citep[e.g.][]{Wang2018, Chainakun2020}. However, \src{} is a source with particular properties~\citep[see][]{Belloni2000}. It is a persistent source that spends decades accreting at near-Eddington rate. The accretion dynamics is more complicated if the disk is thick. For instance, in the case of Polish doughnut model~\citep{Abramowicz1978}, the inner edge of the disk is determined by the fluid angular momentum. If the fluid angular momentum increases as the mass accretion rate increases, the inner edge will move to larger radii. Moreover, when calculating the evaporation of ion-irradiated disks, \cite{Dullemond2005} also find that the inner radius of the cold disk increases with increasing mass accretion rate, which is due to the the larger circumference needed to transfer all the matter. These scenarios may explain what we are seeing on \src{}.}


We would also like to address the potential of our finding on testing general relativity in the strong field regime~\citep[e.g.][]{Bambi2017}. Once the origin of QPOs is understood, the behavior of QPOs can help to reveal the spacetime properties in the vicinity of the black hole~\citep[e.g.][]{Motta2014}. We have been able to test the Kerr hypothesis using the reflection dominated~\citep{Zhang2019} and thermal dominated~\citep{Tripathi2020} X-ray spectra of black hole X-ray binaries, although systematic uncertainties in modelling can somehow affect the analysis~\citep[e.g.][]{Zhang2019a, Liu2019, Riaz2020}. \src{} is a particularly interesting source, in which we can apply continuum fitting method, reflection spectroscopy and QPO modelling together to constrain possible deviation from Kerr metric. We would expect tight constraint from combination of these techniques. This will be further explored in future work.

{\bf Acknowledgments --} The work of H.L and C.B. is supported by the Innovation Program of the Shanghai Municipal Education Commission, Grant No.~2019-01-07-00-07-E00035, the National Natural Science Foundation of China (NSFC), Grant No.~11973019, and Fudan University, Grant No.~JIH1512604. JL thanks supports from the National Natural Science Foundation of China under Grants No. 11733009, U1938103 and U2038101.



\bibliographystyle{apj}


\end{document}